\documentclass[showpacs,epsfig,preprintnumbers,amsmath,amssymb,nofootinbib,20pt]{revtex4-1}
\usepackage{graphicx}
\usepackage{dcolumn}
\usepackage{bm}
\usepackage[all]{xy}
\usepackage{mathrsfs,amsmath,amssymb,amsthm,amsfonts,tikz,graphicx,accents,hyperref, color}

\begin{document}
\hypersetup{linktoc=all,
	colorlinks, linkcolor={palatinateblue},
	citecolor={brightpink}, urlcolor={amaranth}
}
\newcommand{\changeurlcolor}[1]{\hypersetup{urlcolor=#1}}
\def\d{{\rm d}}
\def\Epos{E_{\rm pos}}
\def\ap{\approx}
\def\eff{{\rm eft}}
\def\L{{\cal L}}
\newcommand{\vev}[1]{\langle {#1}\rangle}
\newcommand{\CL}   {C.L.}
\newcommand{\dof}  {d.o.f.}
\newcommand{\eVq}  {\text{EA}^2}
\newcommand{\Sol}  {\textsc{sol}}
\newcommand{\SlKm} {\textsc{sol+kam}}
\newcommand{\Atm}  {\textsc{atm}}
\newcommand{\Chooz}{\textsc{chooz}}
\newcommand{\Dms}  {\Delta m^2_\Sol}
\newcommand{\Dma}  {\Delta m^2_\Atm}
\newcommand{\Dcq}  {\Delta\chi^2}
\newcommand{\nbb}{$\beta\beta_{0\nu}$ }
\newcommand {\be}{\begin{equation}}
\newcommand {\ee}{\end{equation}}
\newcommand {\ba}{\begin{eqnarray}}
\newcommand {\ea}{\end{eqnarray}}
\def\VEV#1{\left\langle #1\right\rangle}
\let\vev\VEV
\def\e6{E(6)}
\def\10{SO(10)}
\def\21{SA(2) $\otimes$ U(1) }
\def\321{$\mathrm{SU(3) \otimes SU(2) \otimes U(1)}$ }
\def\lr{SA(2)$_L \otimes$ SA(2)$_R \otimes$ U(1)}
\def\422{SA(4) $\otimes$ SA(2) $\otimes$ SA(2)}
\newcommand{\AHEP}{%
School of physics, Institute for Research in Fundamental Sciences
(IPM)\\P.O.Box 19395-5531, Tehran, Iran\\

  }
\newcommand{\Tehran}{%
School of physics, Institute for Research in Fundamental Sciences (IPM)
\\
P.O.Box 19395-5531, Tehran, Iran}
\def\roughly#1{\mathrel{\raise.3ex\hbox{$#1$\kern-.75em
      \lower1ex\hbox{$\sim$}}}} \def\lsim{\roughly<}
\def\gsim{\roughly>}
\def\ltap{\raisebox{-.4ex}{\rlap{$\sim$}} \raisebox{.4ex}{$<$}}
\def\gtap{\raisebox{-.4ex}{\rlap{$\sim$}} \raisebox{.4ex}{$>$}}
\def\lsim{\raise0.3ex\hbox{$\;<$\kern-0.75em\raise-1.1ex\hbox{$\sim\;$}}}
\def\gsim{\raise0.3ex\hbox{$\;>$\kern-0.75em\raise-1.1ex\hbox{$\sim\;$}}}

\title{Pico-charged  particles from dark matter decay explain 511 keV line and XENON1T signal}
\date{\today}

\author{Y. Farzan}\email{yasaman@theory.ipm.ac.ir}
\author{M. Rajaee}\email{meshkat.rajaee@ipm.ir}
\affiliation{\Tehran}
\begin{abstract}
There is a robust signal for a 511 keV photon line from the galactic center which may originate from   dark matter particles with masses of a few MeV. To avoid the bounds from delayed recombination and from the absence of the line from dwarf galaxies, in 2017, we have proposed a model  in which dark matter first decays into a pair of intermediate pico-charged particles $C\bar{C}$ with a lifetime much larger than the age of the universe. The galactic magnetic field accumulates the relativistic $C\bar{C}$ that eventually annihilate, producing the  $e^-e^+$  pair that give rise to  the 511 keV line. The relativistic pico-charged $C$ particles can scatter on the electrons inside the direct dark matter search detectors imparting a recoil energy of $E_r \sim$~keV. We show that this model can account for the electron recoil excess recently reported by the XENON1T experiment. Moreover, we show that the XENON1T electron recoil data sets the most stringent bound on the lifetime of the dark matter within this model.
\end{abstract}

\date{\today}
\maketitle
\section{Introduction}

From the cosmological scales down to the galactic scales,  dark matter has demonstrated its existence via gravitational effects. The efforts to discover the particles composing dark matter by direct and indirect dark matter search experiments are ongoing. Although no conclusive discovery has so far been    made,
various observations have been reported that defy an explanation within the standard model and may have a dark matter origin. One of them is the observation of the 511 keV line from the galactic center. Another signal is the recently reported XENON1T electron recoil signal \cite{Aprile:2020tmw}. 

The statistical observation of the 511 keV line is quite robust and its morphology is well reconstructed \cite{Siegert:2015knp}. The shape of this line strongly suggests that it comes from the decay of non-relativistic positronium atoms. The intensity of the line is of course proportional to the density of the positronium atoms. The measured intensity indicates a density for positron in excess of that expected from known sources such as pulsars. The positron excess in the galactic center tantalizingly suggests a dark matter origin. In the scenario proposed in \cite{Boehm:2003bt}, the dark matter pairs of  mass of few MeV annihilate into $e^-e^+$ pairs with a cross section of $10^{-4}$ pb. This simplistic solution is now ruled out because of two reasons: (i) If the annihilation is through the $s$-channel, the $e^-e^+$ production in the early universe would result in the delayed recombination with signatures on the CMB fluctuations that are ruled out \cite{Wilkinson:2016gsy}; (ii) the model predicts a 511 keV photon line from dwarf galaxies but the observations refute this prediction \cite{Sieg}.

In \cite{Farzan:2020llg}, we have proposed a solution that avoids these constraints. In our model, dark matter, $X$, is also a MeVish particle that can be identified with the SLIM particles \cite{Boehm:2006mi,Farzan:2017hol} whose abundance in the universe  is set by the annihilation into the $\nu \nu$ and $\bar\nu \bar\nu$ pairs through the freeze-out scenario. The $X$ particles are metastable with a lifetime larger than
10000 times the age of the universe. In our model, $X$ decays into a pair of $C\bar{C}$ particles which have an electric charge of $q_C\sim 10^{-11}$. With such electric charge, the Larmor radius of these particles in the galaxy will be smaller than the thickness of the galactic disk.
As a result, the $C\bar{C}$ pair will be accumulated in the galaxy despite their velocities being larger than the escape velocity. The $C$ and $\bar{C}$ pairs eventually annihilate and produce  $e^- e^+$ pairs, explaining the 511 keV line. At the recombination, the density of $C$ and $\bar{C}$ would be too small to lead to a significant entropy dump through the $e^-e^+$ production. Moreover, the magnetic fields in the dwarf galaxies are typically too small to accumulate $C\bar{C}$ and lead to a discernible 511 keV line. 

In Ref \cite{Farzan:2020llg}, we had predicted a signal for the electron recoil excess in direct dark matter search experiments.
Recently the XENON1T  detector  has reported an excess of scattered electrons with recoil energies $1-7$ keV over the background. Although one possible solution is the $\beta$ decay of the residue Tritium nuclei in the sample \cite{Aprile:2020tmw} (see also \cite{Bhattacherjee:2020qmv}), this observation has instilled a considerable activity in the field \cite{axion,pp,keVector,boost,downscatter}.
In this paper, we show that our model can simultanously explain the 511 keV line and the XENON1T excess. An alternative solution is proposed in \cite{Ema:2020fit}.

Some of the ideas proposed in the literature to explain the excess include axions \cite{Aprile:2020tmw,axion}, non-standard interaction for the solar pp neutrinos \cite{pp} and the absorption of the background vector dark matter of a mass of a few keV \cite{keVector}. The shape of the electron recoil spectrum cannot be explained by the vanilla WIMP dark matter scattering as the recoil energy off the electrons will be smaller than 1 eV and below the detection threshold. However, boosted dark matter may be able to account for the observed excess \cite{boost}; see, however, \cite{Chigusa:2020bgq}.
Another possibility is down-scattering of inelastic dark matter with a splitting of few keV \cite{downscatter}. 
In our model, the $C$ and $\bar C$ particles wandering in the galactic plane have relativistic velocities. As a result, despite their small mass, the scattering of the $C$ particles off the electrons can impart a sizable recoil energy. In this model, the interaction of the $C$ particles with the electrons (as well as with the protons) 
takes place via the $t$-channel virtual SM photon ({\it i.e.,}  via Coulomb interaction) and  dark photon exchanges. In majority of the unconstrained parameter space, the former dominates so the energy spectrum of the recoiled electron, $dN/dE_r$, is inversely proportional to the square of the recoil energy, $E_r^{-2}$ as expected for Coulomb interaction.
We however show that there is a possibility of cancellation between the contributions from the SM and dark photon exchanges at a given energy bin which provides a better fit to the data.
We also use the XENON1T electron data to derive an upper  bound on the $C$ and $\bar{C}$ abundance ({\it i.e.,} on the fraction of dark matter particles that decay into $C\bar{C}$). 

This paper is organized as follows: In sect. \ref{model}, we present the model. In sect. \ref{BOM}, we review the bounds on the parameters of the  model. In sect. \ref{DDM}, we show how the XENON1T electron recoil excess can be explained within our model and derive a lower bound on the lifetime of dark matter particles. Conclusions are summarized in sect. IV.

\section{The model \label{model}}
The model proposed in Ref \cite{Farzan:2020llg} is based on adding a new $U_X(1)$ gauge symmetry to the Standard Model (SM) gauge group. The new gauge boson has a mixing with the photon both via the kinetic term and via the mass term through Stuckelberg mechanism. This aspect of the model is elaborated on in Ref. \cite{Feldman:2007wj}. Following the notation of \cite{Feldman:2007wj}, the kinetic and Stuckelberg mass terms for the new gauge field $A_\mu$ and the hypercharge gauge field $B_\mu$ can be written as 
\be -\frac{A_{\mu \nu}A^{\mu \nu}}{4}-\frac{B_{\mu \nu}B^{\mu \nu}}{4}-\frac{\delta}{2}A_{\mu \nu}B^{\mu \nu}-\frac{1}{2}\left(\partial_\mu \sigma +M_1 A_\mu +\epsilon M_1 B_\mu\right)^2 \ , \ee
where $\delta, \epsilon\ll 1$. Going to the canonic kinetic and mass basis, we shall have three neutral gauge bosons; {\it i.e.,} the SM $\gamma$ and $Z$ bosons plus a new gauge boson called dark photon, $A'_\mu$. To the linear order in $\epsilon$ and $\delta$, we find that the mass of $A'$ is decoupled from the $Z$ mass and is equal to $M_1$. Neglecting $O(\epsilon^2,\delta^2)$, we find the following  coupling between the SM charged fermions, $f$ and $A'$:
\be q' \bar{f}\gamma^\mu f A'_\mu  \ \ \ \ \ {\rm where} \ \ \ \ \  q'=e \cos \theta_W (\epsilon-\delta)Q_f, \ee
in which $\theta_W$ is the weak (Weinberg) mixing angle.
 Notice that the mass  and couplings of the dark photon can have arbitrary values independent of the SM gauge boson masses.  The intermediate $C$ particles into which the $X$ dark matter particle decays are charged under  the new $U_X(1)$ symmetry. That is their coupling to $A'$ is given by 
 $g_X J_C^\mu A'_\mu$ where $J_C^\mu$ is the current of the $C$ particles. The electric charge of the $C$ particles is 
 $$ q_C=-g_X \epsilon \cos \theta_W.$$   Notice that $q_C$ is suppressed by the mass mixing between the SM photon and the dark photon, $\epsilon$ and can be arbitrarily small.
 As we shall see below, in order to keep the relativistic $C$ particles produced by dark matter decay inside the galactic disk, the electric charge of these $C$ particles, $q_C$, should be of order of or larger than $$ q_C\sim 10^{-11}.$$ 
 
 The $U_X(1)$ gauge symmetry can be identified with
 $L_\mu -L_\tau$. In this case, $A'$ can  decay to $\nu_\mu \bar{\nu}_\mu$ and to $\nu_\tau \bar{\nu}_\tau$.  As we shall discuss in sect. \ref{BOM}, this can  relax some of the bounds from supernova and beam dump experiments. We should however notice that gauging $L_\mu-L_\tau$ is not an essential part of the scenario and in fact, in Ref. \cite{Farzan:2020llg}, we did not identify the $U_X(1)$ gauge symmetry with $L_\mu-L_\tau$. Ref. \cite{Farzan:2020llg}  focuses on the limit of equality of kinetic and mass mixings, $\delta=\epsilon$. In this limit, the coupling of $A'$ to the SM fermions, $q'$, vanishes and in the absence of $A'\to \nu_\mu\bar{\nu}_\mu,\nu_\tau\bar{\nu}_\tau$  the lifetime of $A'$ becomes much larger than the age of the universe; {\it i.e.,} $>7 \times 10^{45}$ years.
 An electric charge of $q_C\sim 10^{-11}$ is too small to bring $C$, $\bar{C}$ and $A'$ into thermodynamical equilibruim with the plasma but it can lead to  a background $A'$ with a density of $n_{A'}/n_\gamma =10^{-4} (q_C/10^{-11})^2$ \cite{Farzan:2020llg}.
 In the limit of vanishing $q'$, we shall therefore have a background of $A'$ with mass of few keV which will act as a subdominant dark matter component as long as $10~e{\rm V}<m_{A'}< 10 ~ {\rm k}e{\rm V}$ with average number density of $\langle n_{A'}\rangle=10^{-4}n_\gamma$. The local density of $A'$ can be then approximated as $n_{A'}|_{\rm local}= \langle n_{A'}\rangle (\rho_{DM}|_{\rm local}/\langle  \rho_{DM}\rangle) $.
 In the limit discussed in \cite{Farzan:2020llg}, since $A'$ does not couple to the electron, no constraint comes from the stellar cooling  consideration. However, in general for an arbitrary ratio of $\delta/\epsilon$, the coupling of $A'$ to the electrons, $q'$, can be nonzero. In this paper, we shall focus on the general case where $q'$ is nonzero and $A'$ is unstable.
 
 In our model, the main component of dark matter is a scalar particle, $X$ which decays into $C\bar{C}$ with a lifetime larger than 10000 times the age of the universe.
 The coupling between $X$ and $C \bar{C}$ is of course too small to bring $A'$ or $C$ to thermodynamical equilibrium with $X$.
 The model presented in Ref. \cite{Farzan:2017hol} embeds the SLIM scenario \cite{Boehm:2006mi} within which $X$ has a mass of few MeV and a Yukawa coupling of $10^{-4}-10^{-3}$ to SM neutrinos and a right-handed Majorana fermion, $N$ with mass $m_X<m_N<10$ MeV. The upper limit on $m_N$ comes from the contribution to neutrino masses. In other words, the SLIM scenario provides a natural mechanism for generating small Majorana mass for active neutrinos. Within this scenario, the $X$ particles in the early universe come to thermodynamical equilibruim with the background neutrinos with $\langle \sigma(X\bar{X} \to \nu\nu, \bar{\nu}\bar{\nu}) v \rangle \sim 1$~pb. From BBN and CMB, a lower bound of 3.7 MeV is set on the mass of $m_X$ \cite{Sabti:2019mhn}. If the $X$ production mechanism is freeze-in or some other mechanism such that $X$ particles never come to the thermal equilibrium with the SM plasma, this lower bound on $m_X$ does not apply. In the present paper, we shall stay agnostic about the production mechanism and shall fix $m_X$ to 10~MeV.  The $X$ particles in the galaxy are non-relativistic so the $C$ and $\bar{C}$ particles produced in the decay of $X$ will have an energy of $m_X/2$.  The Larmor radius of such particles in the local magnetic field is  
 \be  r_L= 100~pc \left( \frac{3\times 10^{-11}}{q_C} \right) \left( \frac{m_X}{10~{\rm MeV}}\right) \left( \frac{ 1~\mu~Gauss}{B}\right), \ee which is smaller than the galactic disk thickness ($\sim 300$ pc) so the $C$ and $\bar{C}$ particles are accumulated. 
  
  The magnetic field in the galactic disk enjoys an approximate azimuthal symmetry \cite{sunAA}. That is the magnetic lines are circles inside the disk centered around the galactic center. The $C$ particles will spiral around the magnetic fields so their distance from the galactic center will not vary more than $\sim r_L\ll $kpc. Considering that the dark matter density also enjoys an azimuthal symmetry, the local densities of $C$ and $\bar{C}$ will be given by 
  $(\rho_{DM}|_{local}/m_X)f$ where $f$ is the fraction of the $X$ particles that have undergone decay.

 As shown in \cite{Chuzhoy:2008zy}, the supernova shock waves pump energy to the charged particles in the galactic disk, enlarging their Larmor radius and eventually repelling them from the galaxy. The rate of this process is $t_{SNW}^{-1}=(100~{\rm Myr})^{-1}$. The reason why SM charged particles such as the electrons still remain inside the disk is that they can lose energy by various mechanisms with a rate larger than
 $t_{SNW}^{-1}$.
 In the model described in \cite{Farzan:2020llg}, the $C$ particles can lose energy by scattering off the background $A'$ particles and can cool down. As pointed out in \cite{Farzan:2020llg}, if $A'$ particles decay, another  subdominant dark matter component can be introduced to play the role of the coolant. Let us denote this particle with $Y$.
 Similarly to \cite{Farzan:2020llg}, we take $10~{\rm eV}<m_Y<10~{\rm keV}$, $n_Y|_{local}=(\rho_{DM}|_{local}/\langle \rho_{DM}\rangle ) \langle n_Y\rangle$. 
 At each collision, the relativistic $Y$ particle with a velocity of $v$ loses an average 
 energy of 
 $$ \Delta E_C=m_Y \left(\frac{E_C}{m_C}\right)^2 v^2.$$ 
 Thus, the cooling time scale $\tau_E$ down to a velocity of $v_f$ can be estimated as \be \tau_E=\int_{m_C(1+v_f^2/2)}^{m_X/2} \frac{dE_C}{\Delta E_C}\frac{1}{\sigma_S vn_Y}\sim \frac{4\pi m_C^3}{g_Y^4 n_Ym_Y}\left( \frac{1}{v_f}-\frac{1}{v_i}\right), \ee
 where $\sigma_S$ is 
 the  scattering cross section which can be written as  $\sigma_S\sim g_Y^4/(4\pi E_C^2)$. For $Y\equiv A'$, $g_Y=g_X$ and for a scalar $Y$, we define $g_Y$ as the square root of the quartic coupling $\bar{Y}Y \bar{C}C$.
 Taking $n_Y m_Y/\rho_{DM}|_{local}\sim 0.1$ and equating $\tau_E$ with 100 Myr, we find
 $$ v_f =0.08 \left(\frac{0.25}{g_Y}\right)^4 \left(\frac{m_C}{3 {\rm MeV}}\right)^3 \frac{0.1 \times \rho_{DM}|_{local}}{n_Ym_Y}.$$
 When the velocity of the $C$ particles reach this value, the energy gain from the supernova shock waves and the energy loss due to the scattering off the $Y$ particles compensate each other. The acceleration due to the supernova shock waves is of course a stochastic phenomenon and the $C$ particles will have a velocity distribution with the mean value of $v_f$. Simulating the velocity distribution of the $C$ particles is beyond the scope of the present paper. We shall take the velocity of the $C$ particles to be collectively equal to $v_f$ so $E_C\simeq m_C(1+v_f^2/2)$. 
 
  The energies of the electrons and positrons from $C\bar{C}$ will be smaller than $m_C/2$ so as long as $m_C<12$ MeV, the bounds from Voyager \cite{Boudaud:2016mos} and inflight annihilation \cite{Beacom:2005qv,Sizun:2006uh} can be readily satisfied.
 
 In summary, the components of the model are the following: (1) The dark matter candidate, $X$ which is a scalar singlet with a mass of $\sim 10$ MeV. (2) Dark photon, $A'$, which is the gauge boson of the new $U(1)$ symmetry and mixes with the photon. The new gauge symmetry can be identified with $L_\mu-L_\tau$ to facilitate invisible decay modes, $A' \to \nu_\mu \bar{\nu}_\mu$ and $A' \to \nu_\tau \bar{\nu}_\tau$ and therefore to relax the bounds from supernova and from the beam dump experiments. (3)
 The scalar $C$ and $\bar{C}$ particles that are charged under the new $U_X(1)$ and therefore obtain a small electric charge, $q_C$ due to the small mixing between the SM photon and the dark photon. In this model, the dark matter decays into $C\bar{C}$ with a lifetime much greater than the age of the Universe. The magnetic field in the galaxy keeps the $C$ and $\bar{C}$ particles with $q_C\sim 10^{-11}$ inside the galaxy. The annihilation of $C \bar{C}$ into $e^-e^+$ explains the 511 keV line from the center of the galaxy. $C$ and $\bar{C}$ can scatter off the electrons in a direct dark matter detector. (4) Background coolant particles, $Y$ with a mass of few keV coupled to $C$ and $\bar{C}$. The role of these coolants is to compensate the energy pump from the supernova shock waves to the $C$ and $\bar{C}$ particles in the galaxy. 
 In the next section, we will discuss the various bounds on the model.
 
 \section{Bounds on the parameters of the model \label{BOM}}

 In this section, we first enumerate the bounds from cosmology and stellar cooling. We then argue how these bounds from supernova can be relaxed by identifying the $U_X(1)$ symmetry with $L_\mu-L_\tau$. We revisit the various bounds after such identification. We then discuss the bounds on the fraction of the dark matter that has decayed into $C\bar{C}$ from various considerations.

 For $A'$ lighter than a few MeV,  $A'$ can contribute as effective relativistic  degrees of freedom, $\delta N_{eff}$, during nucleosynthesis era on which there are strong bounds. Requiring $\delta N_{eff}<1$ then implies $q'<10^{-9} (m_{A'}/{\rm MeV})^{1/2}$.  As shown by Landau and Yang, a spin one particle such as $A'$ cannot decay into a pair of photons \cite{Yang}.
  $A'$ with a mass of keV can however decay into $\gamma \gamma \gamma$ via an electron loop with a rate of $\sim m_{A'}(q'e^3)^2/(100 \pi^3 (16\pi^2)^2)$. 
  For $100~e{\rm V}<m_{A'}<100 ~{\rm k}e{\rm V}$, there are very strong bounds (as strong as $10^{-15}$) on $q'$ from stellar cooling \cite{stellar} which requires at least a  partial cancellation between $\delta$ and $\epsilon$.
  For $m_{A'}\sim$MeV, the strongest bound on $q'$
   is slightly below $10^{-10}$ which comes from supernova cooling \cite{super-10,Dent:2012mx}. This bound can be relaxed if $A'$ has an extra interaction that can trap it inside the supernova core or can result in a decay into the SM particles with a decay length smaller than $\sim 10$~m. 
   
   Let us take a coupling of $g_{\tau-\mu}A_\mu'(\bar{\nu}_\tau \gamma^\mu \nu_\tau-\bar{\nu}_\mu \gamma^\mu \nu_\mu)$ which comes from identifying $U_X(1)$ with  the anomaly free $L_\mu-L_\tau$ symmetry. For $g_{\tau-\mu}>{\rm few}\times 10^{-7} ({\rm MeV}/m_{A'})^{1/2}$, $A'$ can decay inside the supernova core to $\nu_\mu\bar{\nu}_\mu$ and $\nu_\tau \bar{\nu}_\tau$  with decay length much smaller than the supernova core size so the bound on $q'$ can be relaxed. Indeed, $L_\mu-L_\tau$ gauge interaction at one-loop level induces $\delta$ with $\delta=(e g_{\tau-\mu}/12 \pi^2)\log [m_\tau^2/m_\mu^2]\simeq 0.014 g_{\tau-\mu}$ \cite{Kamada:2015era}.
  Taking $g_{\tau-\mu}\sim  10^{-5}$ and $\delta \sim 10^{-7}$, the value of $(q'g_{\tau-\mu})^{1/2}$ will be below the bound from Borexino \cite{Bellini:2011rx,Harnik:2012ni} and GEMMA \cite{Beda:2009kx,Harnik:2012ni}. Moreover, the bounds from the beam dump experiments \cite{stellar} as  well as from supernova cooling are relaxed because of the fast decay of $A'$ to $\nu_\mu \bar{\nu}_\mu$ and
   $\nu_\tau \bar{\nu}_\tau$ pairs. Thus, with $m_{A'}\sim$few MeV, $g_{\tau-\mu}\sim 10^{-5}$ and $\delta \sim 10^{-7}$ (therefore $q' \sim 3 \times 10^{-8}$), all the bounds will be respected. However, neutrinos obtain a new flavor-dependent self interaction with a rate comparable (or even larger than) the weak interaction. This can affect the neutrino emission duration as well as flavor composition of the emitted neutrinos which can be tested in case of future observation of neutrino burst from supernova explosion. At this range of the parameters, when the temperature drops below $m_{A'}$ in the early universe, $A'$ is both in thermodynamical equilibrium with the neutrinos and with the electrons so, unlike dark matter which interact either with the plasma or with $\nu$ \cite{Wilkinson:2016gsy}, the $A'$ decay does not lead to a new contribution to $N_{eff}$.
  
  In the range $100~{\rm keV}<m_{A'}<1~{\rm MeV}$,  the strongest upper bound on $q'$ also comes from supernova cooling  which is about $\sim 5 \times 10^{-11}$.  Again by opening a fast decay mode, $A'\to \nu \bar{\nu}$ with a decay length smaller than $\sim 10$~m, the bound can be relaxed but the same coupling that leads to $A' \to \nu \bar{\nu}$ brings $A'$ to thermal equilibrium 
  with the plasma  in the early universe so at BBN and at neutrino decoupling era, $A'$ will contribute as three bosonic degrees of freedom which is ruled out by the bounds on $N_{eff}$. Let us however consider the case that $A'$ decays outside the supernova core to $\nu \bar{\nu}$. Similarly to the standard picture, the binding energy of the star will be transfered out by neutrinos leading to a neutrino burst such as the one observed in the case of SN1987a so the bounds from simplistic  supernova cooling consideration does not apply. Thus, if $A'$ decays into $\nu \bar{\nu}$ outside the core, $q' \sim 10^{-9}-10^{-10}$ can still be compatible with the observations of neutrinos from SN1987a and with the famous supernova cooling constraint. Taking $m_{A'}\sim 100$ keV, $q' \sim 10^{-9}-10^{-10}$ and $g_{\tau -\mu}\stackrel{<}{\sim}10^{-9}$, the bounds both from supernova cooling and from $N_{eff}$ \cite{Huang:2017egl} can be relaxed.
  In this case, supernova core evolution should be reconsidered which provides a tool to test the model in case neutrinos from another supernova  explosion are observed.
  If $A'$ decay takes place at the outer layers of the star, it can help with shock revival. Notice that  we are introducing new interaction for neutrinos with the electron. However, since $g_{\tau-\mu}q'/m_{A'}^2\ll G_F$ (both for $m_{A'}\sim$~MeV and for $m_{A'}\sim 100$~keV),  the deviation of the interaction rate of the solar neutrinos in our model from the SM prediction is negligible and cannot account for the XENON1T excess \cite{pp}. 

Remember that to prevent the expulsion of the $C$ particles by the supernova shock waves, we introduce a background coolant particle $Y$ in the galaxy.
At each  collision of $C$ and $\bar{C}$ on a coolant $Y$, they lose only a small fraction of  the energy but since $m_Y\ll m_C$, the same recoil energy is enough to make the velocity of $Y$ larger than the escape velocity and repel $Y$ from the galaxy.  Within the time scale of 100 Myr, each $C$ or $\bar{C}$  particle scatters off  the following number of the $Y$ particles 
$$\int \frac{dE_C}{\Delta E_C}\sim 400 (m_C/3~{\rm MeV}) (10~{\rm keV}/m_Y).$$
Let us take the fraction of dark matter particles that decay to be 
$$f=\Gamma_X t_0$$ in which $\Gamma_X$ is the decay rate of $X$ and $t_0 =13$ Gyr 
 is the age of the galaxy.  Each $X$ decay produces a pair of $C \bar{C}$ so the number of the $Y$ particles per unit volume repelled by the $C\bar{C}$ particles during the history of the galaxy can be estimated as
 \be 
 \delta n_Y\sim \frac{t_0}{t_{SNW}} \left( \frac{2 f\rho_{DM}}{m_X} \right)\int \frac{d E_C}{\Delta E_C}.\ee
 Taking $\delta n_Y <n_Y=0.1 \rho_{DM}/m_Y$, we find
 $$ f<10^{-3} \left( \frac{m_X}{10~{\rm MeV}}\right) \left( \frac{m_Y}{10~{\rm keV}}\right)\left( \frac{\rho_Y/\rho_{DM}}{0.1}\right).$$
 As discussed in \cite{Farzan:2020llg}, there is a stronger upper bound on $f$, from the requirement that $C\bar{C}$ in the galaxy do not annihilate away to $A' A'$:
 \be f< 3\times 10^{-4}\left( \frac{m_X}{10~{\rm MeV}}\right) \left( \frac{0.25}{g_X}\right)^4\left( \frac{m_C}{3~{\rm MeV}}\right)^2.\ee

 \section{Predictions for direct dark matter search experiments \label{DDM}}
 As we found out in the previous section, the pico-charged particles, $C$, move around us with a velocity of $v_f\sim 0.08$. The Larmor radius in the vicinity of the Earth is $5\times 10^8~{\rm km}(10^{-11}/q_C)(m_C/3~{\rm MeV})(0.5~{\rm Gauss}/B)(v_f/0.08)$ which is much larger than the Earth radius. That is the magnetic field of the Earth cannot significantly alter the  density of the $C$ particles around the Earth. As discussed in Ref. \cite{Farzan:2020llg}, the local density of the $C$ particles will be 
 $(\rho_{DM}|_{local}/m_X) f$ in which $\rho_{DM}|_{local}\sim 0.4~
 {\rm GeV}/{\rm cm}^3$. These wandering $C$ particles can scatter off the electrons and protons inside the direct dark matter search experiments such as XENON1T.

 By analyzing the relativistic kinematics, it is straightforward to show that the maximum recoil energy of the $C$ particles with velocity of $v_f$ scattering off non-relativistic particles of mass, $m$, can be written as
 \be  E_{max} =2m v_f^2 \frac{m_C^2}{(m+m_C)^2} \label{max}
 \ee
 where we have neglected corrections of $O(v_f^4)$. This maximum recoil corresponds to backward scattering.
  Taking $m_C\sim 1-5$ MeV, $m=m_e$ and $v_f =0.08$, we find $ E_{max} =3-5.5$~keV which is tantalizing in the range of electron recoil excess observed by XENON1T \cite{Aprile:2020tmw}. Taking $m$ equal to the mass of  Xenon or Argon, we find $ E_{max} $ to  be much smaller than 1~eV so the bounds on  the scattering off nucleons from direct search experiments such as XENON1T \cite{Aprile:2018dbl} and DarkSide \cite{DarkSide-50} can be satisfied.
 The CRESST detector is made of CaWO$_4$ and aims at a detection threshold of 100 eV but $E_{max}$ scattering off even the Oxygen nucleus will be one order of magnitude below this state-of-the-art
 detection threshold \cite{Schieck:2016nrp}.

 The spectrum of recoiled electrons from a unit of  mass of the detector can be estimated as 
 $$\frac{dN}{dE_r}=\frac{Z_{out}}{m_N}(2 f\frac{\rho_X}{m_X}) \int f_C (v) \frac{d\sigma}{d E_r}vdv$$
 where $m_N$ is the mass of the nuclei composing the detector. For Xenon, $m_N=131$~GeV. $Z_{out}$ is the number of the electrons per nucleus with binding energy smaller than the recoil energy. As shown in \cite{Hsieh:2019hug}, taking step function of the binding energy for the cross section is a valid approximation. For Xenon, we take $Z_{out}=44$ which is the number of the electrons in the outer orbitals with principal quantum number $n=3,4,5$. For these electrons, unlike the electrons in the inner orbitals, we can neglect the electron velocity. $f_C(v)$ gives the velocity distribution of $C$ and $\bar{C}$ particles. As discussed before, we shall take 
 $f_C(v)=\delta( v-v_f)$ for simplicity.  As far as $|\epsilon|/2m_eE_r\gg |\epsilon-\delta|/(2m_eE_r+m_{A'}^2)$, the $t$-channel photon exchange gives the dominant contribution to the $C$ particles scattering off  the electrons.  The differential cross section of scattering of the $C$ and $\bar{C}$ particles with electric charge of $q_C$ off the electrons can be written as \cite{Harnik:2019zee}
 \be \label{diffSigma}
 \frac{d\sigma}{dE_r}=\frac{e^2 q_C^2 }{8 \pi m_e v^2}\frac{1}{E_r^2} ~~ {\rm where}~~ E_r<E_{max}=2 m_e v_f^2\left(\frac{m_C}{m_C+m_e}\right)^2 .\ee 
 Notice that at $E_r=E_{max}$, $d\sigma /dE_r$ is nonzero.
 For general values of $\delta/\epsilon$, the coupling of the dark photon, $A'$, to the electron, $q'$ is nonzero so the dark photon exchange between the electron and $C$ also contributes to the scattering amplitude. Thus, $d\sigma/dE_r$ should be modified by an extra factor of 
 \be \left( 1+ \frac{2 m_e E_r}{2 m_e E_r+m_{A'}^2}\frac{\delta -\epsilon}{\epsilon}\right)^2 \label{modi} .\ee
 Of course, in the limit $\epsilon \to \delta$ that has been explored in Ref. \cite{Farzan:2020llg}, the second term vanishes and we recover Eq.~(\ref{diffSigma}). In fact, for $m_{A'}\sim 1-100$~keV (for $m_{A'}\sim$MeV$-50$~MeV), the bounds on $q'$ from stellar cooling (supernova cooling) render the second term negligible implying that the photon exchange is the dominant one. For the time being, let us neglect the correction due to $A'$ exchange and analyze whether the XENON1T electron recoil data can be explained by the photon exchange. We shall return to the correction in Eq. (\ref{modi}) later.
  Neglecting the contribution from the dark photon exchange,  we  find
 \be \frac{dN}{dE_r}=\frac{1225}{ton.year. keV} \left(\frac{2~{\rm keV}}{E_r}\right)^2\left(\frac{q_C}{10^{-11}}\right)^2\frac{f}{10^{-4}} \frac{10~{\rm MeV}}{m_X}\frac{0.08}{v_f}\Theta\left(E_{max}- E_r\right) .\label{thet}\ee

\begin{figure}[t]
	\includegraphics[scale=0.55]{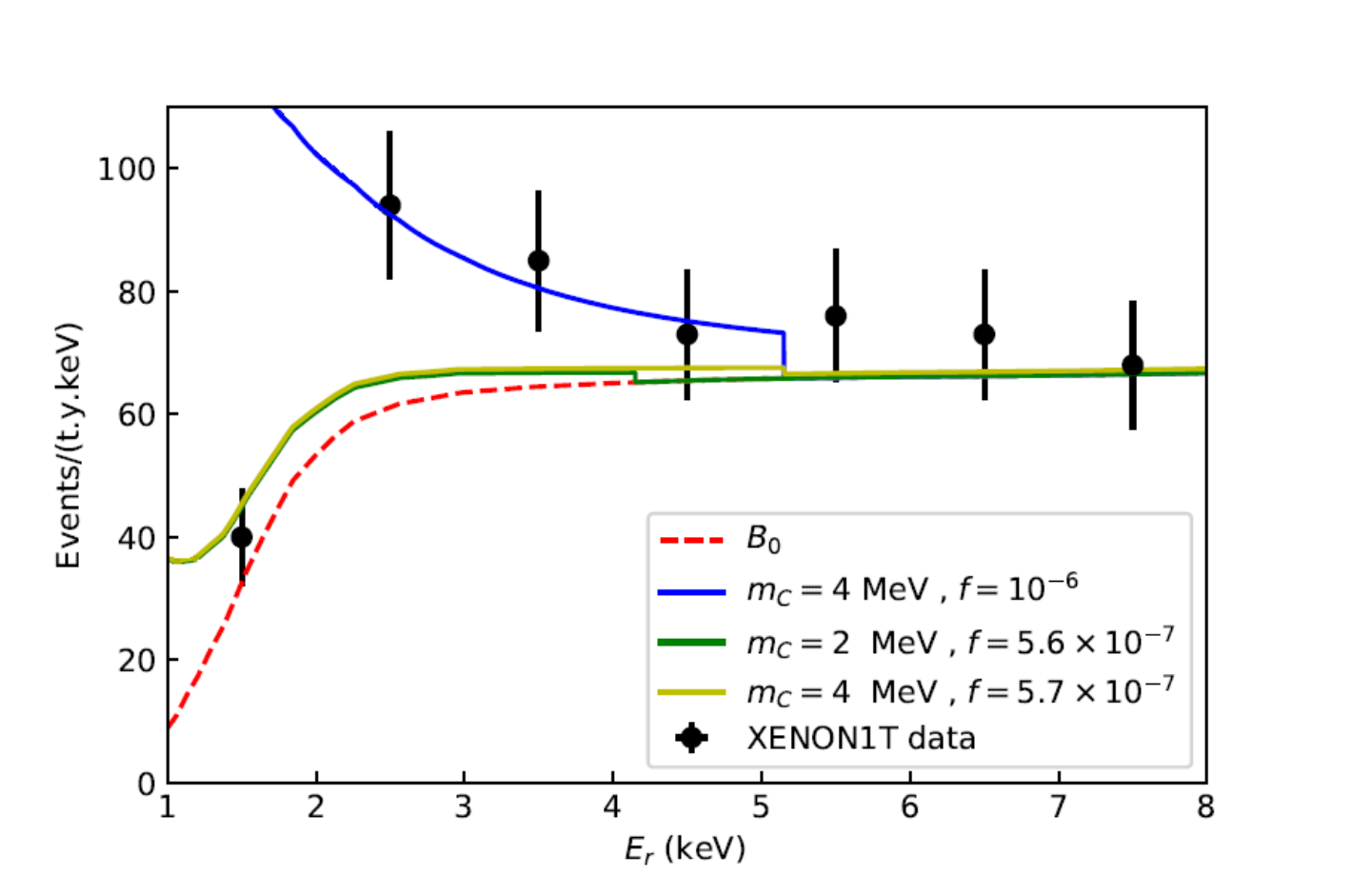}
	\caption{ XENON1T electron recoil data compared with the prediction of our model without considering the contribution from the $A^\prime$  exchange (Eq (\ref{modi})).  The black dots represent the XENON1T data with
		their experimental errors shown by the vertical bars  \cite{Aprile:2020tmw}. The blue   curve  indicates the  expected   signal plus background  plotted for $m_C= 4$ MeV and for our best fit point of  $f=4 \times 10^{-6}$, excluding the first bin (fitting the bins with $2~{\rm keV}<E_r< 8~{\rm keV}$). The green curve  indicates the predicted signal plus background for $m_C= 2$ MeV  ({\it i.e.,} $E_{max}=4$ keV),  taking the corresponding best fit for the first seven bins: $f=5.6 \times 10^{-7}$. The olive curve shows the same for $m_C= 4$ MeV ({\it i.e.,} $E_{max}=5$ keV)    and corresponding best fit $f=5.7 \times 10^{-7}$.  The red dashed curve shows the background. 
	} \label{fit}
\end{figure}

\begin{figure}[t]

\includegraphics[scale=0.7]{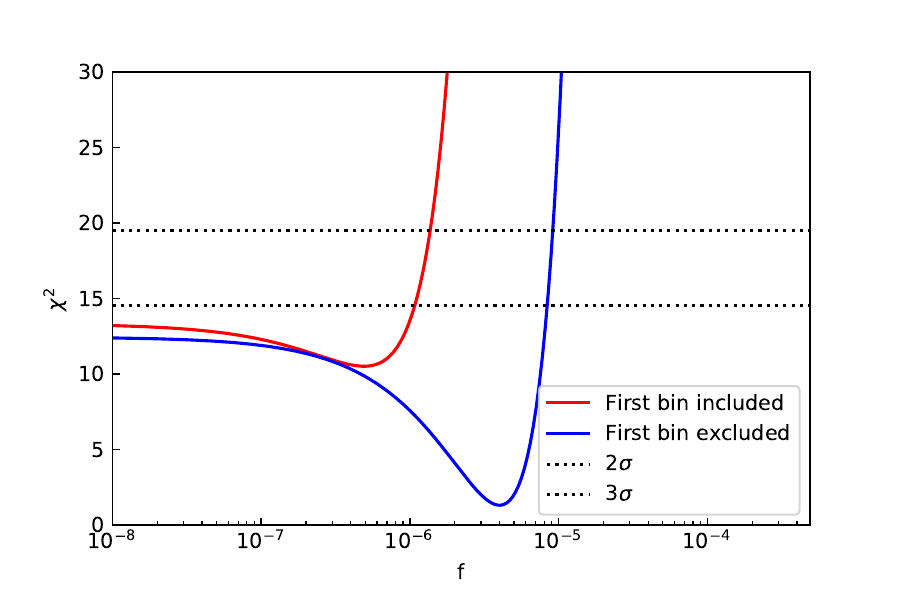}
\caption{ $\chi^2$ vs. $f$.  The red  curve  shows   $\chi^2$ for the first seven bins with $1~{\rm keV}<E_r< 8~{\rm keV}$ as a function of $f$. 
   The minimum of $\chi^2$ lies at   at $f=5.7 \times 10^{-7}$ and is aqual to $10.1$. The blue  curve demonstrates   $\chi^2$  computed excluding  the first bin ({\it i.e.,} considering only the bins with $2~{\rm keV}<E_r< 8~{\rm keV}$).  The best point fit is located at $f=4 \times 10^{-6}$ and is aqual to  $1.4 $. We have taken    $m_C\sim 4$ MeV to draw these curves but they appear to be very robust against varying $m_C$.  The black  dotted lines indicate the $2\sigma$ and $3\sigma$ limits.  } \label{chi}
\end{figure}
 \begin{figure}[t]
\includegraphics[scale=0.85]{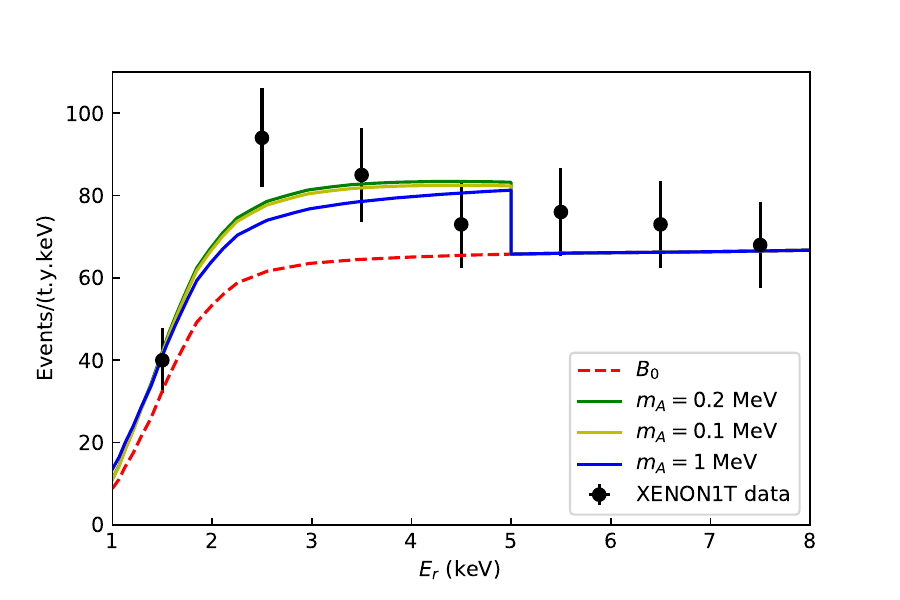}
\caption{ XENON1T electron recoil data compared with the prediction of our model    taking into account the contribution from the $A^\prime$  exchange (Eq (\ref{modi})). 
The blue   curve  shows  the signal plus background  assuming $m_{A^\prime} > 1$ MeV  and taking the best fit values,  $f=10^{-7}$ and $(\delta -\epsilon)/\epsilon=-2 \times 10^3 (m_{A'}/{\rm MeV})^2$.
The green   (olive) curve  shows  the signal plus background assuming $m_{A^\prime} = 0.1$ MeV  ($m_{A^\prime} = 0.2$ MeV) and taking  the  best fit values:  $(\delta -\epsilon)/\epsilon=-13.9 $ and $ ~ f=6.6\times 10^{-7}$ ($(\delta -\epsilon)/\epsilon=-76$ and $ f=2\times 10^{-7}$). We have taken   $m_C= 4$ MeV which corresponds to $E_{max}=5$ keV. The red dashed curve shows the background. 
  } \label{fit2}
\end{figure}

Let us now   analyze the  	XENON1T data  within the framework of our model.  Throughout our analysis we take $v_f=0.08$, $q_C=10^{-11}$ and $m_X=10$~MeV. We  use the binned data shown in Fig 4 of \cite{Aprile:2020tmw}  and  define $\chi^2$ as follows
\begin{equation} \label{chi2}
\chi^2=
\sum_{bins}\frac{[N^{pred} _i - N^{obs} _i]^2 }{\sigma_i^2}
\end{equation}
where $N^{obs}_i$ is the  number of the observed events at each bin and $N^{pred}_i$ is the predicted number of events which is the sum of the   background and the signal from the $C$ scattering in the $i$th bin. The values of $\sigma_i$ (the uncertainty), $N^{obs}_i$ and the background at each bin are extracted from Fig 4 of \cite{Aprile:2020tmw}.
Since the reported excess shows up at $E_r<8$~keV, we only consider the first seven bins. Notice that because of the theta function in Eq. (\ref{thet}), for higher energy bins, $N_i^{pred}$ is equal to the background. From Eqs. (\ref{max},\ref{thet}), we observe that the dependence on $m_C$ is mild. For
$m_C=4$~MeV ($m_C=2$~MeV), the minimum of $\chi^2$ for the first seven bins lies at $f=5.7\times 10^{-7}$ ($f=5.6 \times 10^{-7}$) and is $\chi^2=10.1$ ($\chi^2=10.3$). The improvement relative to the case of $f=0$ is mild and this is because while the signal is predicted to increase at low energies as $E_r^{-2}$  there is no  deviation from the background in the first bin at $E_r=1$~keV.
Fig \ref{fit} shows our prediction for $m_C=2$~MeV and $m_C=4$~MeV with corresponding best fits for $f$.
As seen from Fig. \ref{fit}, the scattering of the $C$ particles off the electrons with its $E_r^{-2}$ behavior provides a good fit for low energy bins except the lowest energy bin which is dangerously  close to the detection threshold of XENON1T. As discussed in \cite{Aprile:2020tmw}, the background from $^{214}$Pb has also large uncertainty at this bin. Excluding this data point,  $f=4\times 10^{-6}$ will provide the best fit with $\chi^2=1.4$   with 5=6-1 degrees of freedom. This fit is also shown in Fig \ref{fit} with $m_C=4$ MeV. As expected, there are jumps in the curves at maximum recoil energies 4~keV and 5~keV corresponding to the backward scattering of the $C$ particles. At these recoil energies, the scattering amplitude is nonzero so a jump in the curves or equivalently a Heaviside $\theta$-function in Eq. (\ref{thet}) is expected.

Fig.~\ref{chi} shows  $\chi^2$ vs. $f$ for the bins with $E_r<8$~keV,  including and excluding the first bin. To draw this figure we have taken $m_C=4$~MeV, but varying the value of $m_C$ does not considerably change the curves. Notice that for all values of $f$, the red line which includes the  first data point is above the blue line. For $f<10^{-7}$, this is simply because  when we exclude the first bin, one (positive) term in computation of $\chi^2$ is removed so it becomes smaller.   The  horizontal lines show the limits for 2$\sigma$ and $3\sigma$. Thus, taking all the bins with $E_r<8$~keV at the face value, we find  an upper bound of $1.5 \times 10^{-6}$ on $f$ at 3~$\sigma$. Excluding the first energy bin of the XENON1T electron excess, the bound relaxes to $10^{-5}$. Notice that these are the strongest bounds on $f$ so far. Remembering that $f=\Gamma_X t_0$, the upper bound on $f$ can be interpreted as a lower bound on the $X$ lifetime. That is we have found that the lifetime of $X$ should be larger than $10^5-10^6$ times the age of the Universe. The natural question that arises is that whether with this stringent bound, the model can still explain the 511~keV line. 	
To account for the 511 keV line with $f=O( 10^{-7})$, as shown in Ref. \cite{Farzan:2020llg}, the annihilation cross section of $C\bar{C}$ should be $10^{-4} ~{\rm b} ( 10^{-7}/f)^2 (m_X/10~{\rm MeV})$. 
The $C\bar{C}$ pair first annihilate to the intermediate  $\phi$ particles which eventually decay into $e^-e^+$ pairs \cite{Farzan:2020llg}. To have such annihilation, we may introduce quartic coupling $\lambda_{\phi C}|\phi|^2 |C|^2$. An annihilation cross section of 0.1~mb can be obtained with $\lambda_{\phi C}\sim 6 \times 10^{-3}$.

 Let us now consider the correction in Eq.~(\ref{modi}) and check whether by considering the contribution from the $A'$ exchange, the first data point can also be fitted. As we discussed in sect. \ref{BOM}, for $100~{\rm keV}<m_{A'}< $few MeV, the bounds on $q'$ ({\it i.e.,} on $\delta -\epsilon$) from supernova cooling can be relaxed by introducing new interactions between $A'$ and neutrinos. 
 Let us first consider the range $m_{A'}\sim$few MeV. 
	Within this range, $2m_e E_r\ll m_{A'}^2$ so the modification factor in Eq.~(\ref{modi}) can be approximated as $(1+(2m_e E_r/m_{A'}^2)(\delta -\epsilon)/\epsilon)^2$. Thus, we can take $f$ and $(\delta -\epsilon)/(\epsilon m_{A'}^2)$ as two free parameters to fit the data. Taking $m_C=4$~ MeV, the best fit for these parameters turns out to be $f=10^{-7}$ and $(\delta -\epsilon)/\epsilon=-2 \times 10^3 (m_{A'}/{\rm MeV})^2$ which for $\epsilon=-q_C/(g_X\cos \theta_W)\sim 10^{-11}/(g_X\cos \theta_W)$ corresponds to $q' \sim (10^{-8}/g_X) (m_{A'}/{\rm MeV})^2$. 	As discussed in the previous section, turning on the interaction with neutrinos such that $A'$ can decay with a decay length smaller than 10~m into $\nu \bar{\nu}$, the bounds from supernova cooling can be relaxed, ruling in this value of $q'$. The minimum $\chi^2$ is 4.7 for $5=7-2$ degrees freedom which corresponds to a goodness of fit of 45~\% (one-sided). 
	This significant improvement in the fit is due to a cancellation between the contributions from the dark photon and the SM photon exchange $t$-channel diagrams in the first bin. Fig. \ref{fit2} shows this fit.
	
	Let us now focus on the range $m_{A'}\sim 100$ keV.  Notice that in this range, $2 m_e E_r$ in the denominator of Eq. (\ref{modi}) cannot be neglected and we should use the whole formula to fit the data. Taking $m_C=4$ MeV and $m_{A'}=100$~keV ($m_{A'}=200$~keV), the best fit will correspond to
	$(\delta -\epsilon)/\epsilon=-13.9$ and $f=6.6\times 10^{-7}$ ($(\delta -\epsilon)/\epsilon=-76$ and $f=2\times 10^{-7}$) and to the minimum $\chi^2=4.11$ ($\chi^2=4.5$) which is again an excellent fit thanks to the cancellation at the first bin. 
	As Fig. \ref{fit2} demonstrates   the predictions corresponding to these fits are very close to each other. 
	Notice that for such values of $\delta$, $q'$ will be just slightly above the bound from supernova: $q'\sim 10^{-10}-10^{-9}$.
	As discussed in the previous section, these values of $\delta$ (or $q'$) are allowed provided that there is
	a small coupling between $A'$ and neutrinos such that $A'$ produced inside the supernova core can decay outside the core to neutrino pairs. This can lead to a testable effect in the flavor, energy spectrum and the emission duration of neutrinos from the supernova explosion.
	Moreover, through the $q'$ coupling, the $A'$ particles can be produced in the early universe and contribute to $\delta N_{eff}$ with an  amount  of $<1$. Probing smaller values of $\delta N_{eff}$  can test this model with $q'\sim 10^{-9}-10^{-10}$.


\section{Summary and conclusions \label{summary}}

We have examined whether the electron recoil excess recently reported by the XENON1T collaboration can be explained by the electromagnetic interaction of relativistic pico-charged particles, $C$ and $\bar{C}$ produced in the decay of relatively light dark matter particles with a mass of $\sim 10$~MeV.
This scenario was originally proposed in \cite{Farzan:2020llg} to account for the 511 keV line coming from the galactic center. The magnetic field of the galaxy keeps the $C$ and $\bar{C}$ pico-charged particles inside the galaxy. The $C$ and $\bar{C}$ particles can interact with the electrons and protons inside the detector.  The recoil energy imparted on the nuclei will be much smaller than the detection threshold but that imparted on the electrons can be around few keV and in the detectable range for XENON1T.

The $C$ particles, having an electric charge of $q_C$, interact with the electron via $t$-channel photon exchange. If the dominant interaction mode is this virtual photon exchange, the dependence of the scattering will have  an interaction of  form $E_r^{-2}$ which is characteristic of the Coulomb interaction. The $E_r^{-2}$ dependence provides an excellent fit to the events with $2~{
\rm keV}<E_r<8~{\rm keV}$; however, the events in the first bin with $1~{\rm keV}<E_r<2$ keV are compatible with the background and deviate from this $E_r^{-2}$ behavior of the signal, calling for a cancellation. The underlying model for pico-charged particles is based on a new $U_X(1)$ gauge symmetry with new gauge boson called dark photon ($A'$) mixed with the SM photon. The pico-charged particles are charged under the new $U_X(1)$  so they couple  to the dark photon.  $A'$ can have a small coupling with the electron, $q'$. The values of $q_C$ and $q'$ are respectively given  by the kinetic and mass mixing between the SM and dark photons. Thus, $q_C$ and $q'$ can have independent values.  The $C$ particles can interact with the electron also via dark photon exchange. If $|q_Ce/(2E_r m_e)|$ and  $|q'g_X/(2E_r m_e+m_{A'}^2)|$ are of the same order, there is a possibility of cancellation in the first recoil energy bin. For $m_{A'}<100$ keV, the strong bounds from stellar cooling imply $q' g_X\ll q_C e$ but we find that  with $100~{\rm keV}<m_{A'}<{\rm few ~ MeV}$ and with appropriate choice of $g_Xq'/q_C$, our model provides an excellent fit to the low recoil energy excess. Such values of $q'$ lie slightly above the supernova cooling bounds but we have argued that if $A'$ can decay into neutrino pairs, the bounds from supernova cooling can be relaxed. Still the $A'$ production would leave its imprint in supernova neutrino flavor and energy spectra as well as in the neutrino emission duration which  provides an alternative method to test the model in case of observing a supernova neutrino burst in future.
$A'$ can be produced via the $q'$ coupling in the early universe, providing a  contribution to $N_{eff}$ of  order of $0.1$ or smaller. As a result, the model can also be  tested by better determination of $N_{eff}$ through studying CMB and/or BBN.  The distinct energy dependence of the electron recoil excess prediction in this model can be tested by further and more precise data from direct dark matter search experiments.

In order to accumulate the $C$ and $\bar{C}$ particles in the galactic disk, the electric charge of $C$ should be larger than $10^{-11}$. The strongest upper bound on the electric charge comes from supernova cooling which is $\sim 10^{-9}$ \cite{Davidson:2000hf}. The sensitivity of XENON1T electron recoil energy spectrum to the mass of $C$ is mild. As long as only the electron recoil data of XENON1T is concerned, there is no upper bound on $m_C$ provided that $m_C<m_X/2$. However, if we also want to  explain the 511 keV excess, the annihilation of $C\bar{C}$ should produce $e^- e^+$ pairs. The energy of the  $e^-$ and $e^+$  pair at injection increases with increacing the $C$ mass. Since there are bounds on the energy of $e^+$ at injection from Voyager \cite{Boudaud:2016mos} and from inflight annihilation \cite{Beacom:2005qv,Sizun:2006uh}, $C$ should be lighter than $\sim 10$~MeV. For such light $C$, the recoil energy of the scattering off nuclei will be smaller than the sensitivity threshold of direct dark matter search experiments so the bounds from CRESST or DarkSide do not apply. To account for the electron recoil excess with  $q_C \sim 10^{-11}$, the fraction of dark matter particles decaying into $C$ and $\bar{C}$, $f =\Gamma_X t_0$, should be few$\times 10^{-7}-10^{-6}$ which readily satisfies the various bounds on the decay of dark matter to relativistic particles \cite{Farzan:2020llg,Audren:2014bca}. As shown in Fig. \ref{fit} dependence on $m_C$ is mild as expected.  For $m_C\gg 5$~MeV, there are however constraints from the null results on the scattering off the nuclei in the direct dark matter search experiments.

We have shown that the XENON1T electron recoil data provides an upper bound of $10^{-6}$ on $f$ at 3~$\sigma$. This is the strongest bound on $f$ so far. In other words, from the XENON1T data we have found  that the lifetime of the dark matter particles decaying into $C\bar{C}$ should be larger than $10^6$ times the age of the Universe.
We have  shown that with such small values of $f$, it is still possible to account for the 511 keV line. As discussed in \cite{Farzan:2020llg}, this solution to the 511 keV line can be tested by studying the correlation between the magnetic field in the dwarf galaxies and the intensity of the 511 keV line emitted by them. 
\subsection*{Acknowledgments}

This project has received partial funding from the European Union's Horizon 2020 research and innovation programme under the Marie Sklodowska-Curie grant agreement No. 690575 (RISE InvisiblesPlus) and No. 674896 (ITN Elusives) and the European Research Council under ERC Grant NuMass (FP7-IDEAS-ERC ERC-CG 617143). YF has received partial financial support from Saramadan under contract No.~ISEF/M/99169. YF is grateful to the  ICTP staff and associate office for partial financial support and warm hospitality. MR would like  to thank Pouya Bakhti for useful discussions.



\end{document}